# An Efficient Retransmission Based on Network Coding with Unicast Flows

Zhiheng Zhou, Liang Zhou, Yuanquan Tan, Xing Wang
National Key Laboratory of Communication
University of Electronic Science and Tech of China
Chengdu, China
E-mail: {zhzhou, lzhou}@uestc.edu.cn

*Abstract*—Recently, network coding technique has emerged as a promising approach that supports reliable transmission over wireless loss channels. In existing protocols where users have no interest in considering the encoded packets they had over coding or decoding operations, this rule is inefficient and expensive. This paper studies the impact of encoded packets in the reliable unicast network coding via some theoretical analysis. Using our approach, receivers do not only store the encoded packets they overheard, but also report these information to their neighbors, such that users enable to take account of encoded packets in their encoding procedures as well as decoding operations. Furthermore, we present a redistribution algorithm to maximize the coding opportunities, which achieves better retransmission efficiency. Finally, theoretical analysis and simulation results for a wheel network illustrate the improvement in retransmissions efficiency due to the encoded packets.

*Keywords*--Wireless Network; Network Coding; Unicast; Retransmission;

## I. INTRODUCTION

It is well-known that wireless networks are error-prone by reason of fading and interference. Many researches have revealed that IEEE 802.11-based wireless networks suffer from severe noise, which causes the failure of transmission, in wide conditions [1]. Hence, automatic repeat request (ARQ) [2] and Hybrid ARQ (HARQ) technique are used to make a wireless link reliable and further improve end-to-end throughput.

Network coding [4] is a recently introduced paradigm to increase the network bandwidth efficiency and reliability, in respect that it enables to mix multiple incoming packets in a single transmission rather than just forward these incoming packets to output links one by one, which disaccustom the traditional store-and-forward way [3]. As an laudable network coding paradigm, COPE [5] proposed a practical XOR-based network coding protocol with the concept of opportunistic coding and opportunistic listening for intersection flows in wireless networks. It applied cumulative ACK in its COPE layer to provide reliable transmission. After that, reliable transmission based on network coding for wireless network has attracted much attention because it has the potential to cope with unreliable wireless links. D. Nguyen et al. [7] and Rozner et al. [8] first utilized network coding to provide reliable retransmission over the unicast and multicast networks. Following [7], Tran et al. [9] incorporated network coding into FEC technique to increase the bandwidth efficiency of wireless single-hop reliable transmission. CLONE [10], which is a generalization of COPE, took into account lossy links and introduced redundancy to deal with packet-loss, where nodes will receive several coded packets and they will be able to decode in more than one way the packet which is of interest to them. Authors of [11] presented a novel MAC protocol named Phoenix, which can leverage network coding method and cooperative relaying techniques to improve ARQ in wireless networks. Furthermore, a lot of researchers have studied network coding for reliable transmission. [12-14].

However, in existing protocols where users have no interest in considering the encoded packets they had in coding or decoding operations, this rule is expensive and inefficient. Using the conventional network coding protocols, when users overheard encoded packets, they just discard them or simply store them but don't report these information to any neighbors. Consequently, when the users make coding decisions, they cannot explore coding opportunities from those encoded packets.

Let us take an example in Fig. 1 to illustrate how to explore coding opportunities from encoded packets.

Fig. 1(a) depicts a wheel topology, where node C is surrounded by six nodes evenly distributed along the cycle. There are three flows denoted by $f_1(S_1 \rightarrow R_1)$, $f_2(S_2 \rightarrow R_2)$, and $f_3(S_3 \rightarrow R_3)$. We use dotted lines --- to denote the overhearing links, and use square brackets [ ] to denote the packet(s) overheard by nodes. In this example, node *C* has 3 packets in its output queue, $P_1$, $P_2$ and $P_3$, which came from $S_1$, $S_2$ and $S_3$, respectively. The destination of each packet in C's queue is listed in Fig. 1(b). We assume that $R_1$ and $R_2$ have correctly overheard $[P_2]$ and $[P_1]$ severally. Hence, node *C* can broadcast the encoded packets $P_1 \oplus P_2$ to both node $R_1$ and $R_2$, where the operator "$\oplus$" between the two packet notations is equivalent to XOR in $GF(2)$. And $P_3$ have to be unicast to $R_3$ alone in the original transmissions, for the reason that none of the receivers have enough information to decode a coded packet, who contains $P_3$ (e.g. $P_1 \oplus P_2 \oplus P_3$).

Due to the broadcast and error-prone nature of wireless channels, when *C* transmitted $P_1 \oplus P_2$ and $P_3$ separately, $R_1$, $R_2$ and $R_3$ may lose their intended packet but overhear the other one. Table in Fig. 1(c) shows five possible packet-loss states, where at least one of nodes $R_1$ and $R_2$ lost $P_1 \oplus P_2$ but overheard $P_3$, meanwhile $R_3$ lost $P_3$ but overheard $P_1 \oplus P_2$. We assume feedback is reliable and immediate. Using the traditional network coding, $R_3$ would discard $P_1 \oplus P_2$ or simply store it but not report to *C*. Wherefore, node *C* is not able to retransmit $P_1 \oplus P_2 \oplus P_3$, since it "knew" that $R_3$ did not enable to abstract its corresponding native packet from this encoded packet, i.e. node *C* would have to retransmit the lost packets in two time slots. However, under these cases, if $R_3$ stored $P_1 \oplus P_2$ and notified *C*, the best retransmission decision could be made, where node *C* would be to send out $P_1 \oplus P_2 \oplus P_3$, which would allow all three receivers to receive their intended packets all at once. For instance, with the first loss state, $R_2$ could extract $P_2$ by XORing $P_1$ and $P_3$ with $P_1 \oplus P_2 \oplus P_3$, and $R_3$ could XOR $P_1 \oplus P_2$ with $P_1 \oplus P_2 \oplus P_3$ to obtain $P_3$. In this way, we would save one transmission in comparison with those conventional approaches, resulting in the improvement of retransmission efficiency.

The above discussion illustrates the benefit of encoded packets and also raises some interesting questions and challenges. To solve these issues, we propose an efficient retransmission approach based on XOR-network coding in this paper. Our contribution is summarized as follows:
- We introduce a system model of the wheel topology, and then study a theoretical analysis in terms of number of retransmissions with or without considering the encoded packets.
- We formally define the *weight of packet-loss pattern*, which helps us to evaluate the performance of the proposed approach more precisely.
- We present a redistribution algorithm to maximize the coding opportunities, which achieves better retransmission efficiency.

The remainder of this paper is organized as follows. In Section II we describe a system model in the context of wireless wheel scenario. In Section III, we present some theoretical derivations on the total number of retransmissions with the traditional and the proposed schemes. In Section IV, theory and simulation results are illustrated. Finally, conclusion is given in Section V.

## II. SYSTEM MODEL AND ASSUMPTIONS

If multiple flows traverse the same node, we called such node *coding node* (unless otherwise stated we simply call it node *C*) and there is the opportunity to apply network coding techniques to improve the overall retransmission efficiency for each flow crossed it. Consequently, we only consider the retransmissions between node *C* and its neighbors even for the wheel scenario in this paper. And we also consider that the throughput rate of each flow over the networks has already stabilized. Furthermore, we assume node *C* employs a sufficient large retransmission buffer to avoid too early rescue process. In particular, each receiver would request a distinct set of loss packets, which from node *C*'s point of view, corresponds to supporting different unicast sessions. Hence, we introduce the following assumptions of the wireless unicast retransmission model.
- There are $N$ ($N > 2$) receivers $R_i$ ($1 \leqslant i \leqslant N$), and the coding node retransmits lost packets after a fixed time slot $\Delta T$.
- We suppose that route of each flow would pass through the coding node to maximize coding

opportunities [6].
- Node *C* can always know the current packet-loss states of both native and encoded packets at all receivers. This can be carried through by using positive acknowledgements (ACKs).
- To simplify the analysis, we assume all the feedback are instantaneous and reliable.
- Packet lost rates between node *C* and each receiver are mutually independent and follow the Bernoulli distribution, where each packet is lost with a fixed probability $\omega_i$ (*i* is the receiver's ID, $1 \leqslant i \leqslant N$) at each receiver.

According to the example as shown in Fig. 1, we notice that the benefit of encoded packets is emerged, because node *C* cannot mix flow $f_3$ with $f_1$ and/or $f_2$, but can combine $f_1$ and $f_2$ in original transmissions, since $R_3$ cannot listen to $S_1$ and $S_2$, and $R_1$ and $R_2$ can overhear $S_2$ and $S_1$ respectively. It leads to the following definition for irrelevant flow and relevant flow that will be used to help us to evaluate and depict the existing and proposed network coding schemes.

**Definition 1.** *We call a flow an irrelevant flow, if the coding node is unable to combine it with any other flows in original transmissions; otherwise we call it a relevant flow.*

With this definition, flow $f_1$ and $f_2$ are relevant flows and $f_3$ is an irrelevant flow in Fig. 1(a). Let us consider the wheel topology, which consists of *N* receivers. Thereby, if each client requests *K* distinct packets, then node *C* needs to successfully deliver a total of $K \times N$ packets to all of them. To plainly represent the packet-loss state, we define packet-loss pattern as follow:

**Definition 2.** *Packet-loss pattern $\rho_P$ is a row vector that represents the current packet-loss state of packet P at all the receivers, thus its dimension is equal to the number of the receivers. When packet P is successfully obtained by receiver $R_i$, the $i^{th}$ entry in $\rho_P$ will be marked 1, or else 0.*

In this paper, the packet-loss pattern relating to the native packets is simply called *native pattern*, and *coded pattern* denotes the one relating to the encoding packets. The following definitions play crucial roles in this paper.

**Definition 3.** *The weight of a packet-loss pattern $W(\rho)$ is the number of non-zero elements in $\rho$. In particular, $W(\rho) < N$ for N-receiver scenarios.*

For instance, the coded pattern $\rho_{P_1 \oplus P_2}$ relating to the second scenario (the second row) as shown in Fig. 1(c) is $\rho_{P_1 \oplus P_2} = \begin{bmatrix} 1 & 0 & 1 \end{bmatrix}$ and the weight of $\rho_{P_1 \oplus P_2}$ is $W(\rho_{P_1 \oplus P_2}) = 2$, which denote that all the receivers correctly obtained packet $P_1 + P_2$ aside from node $R_2$.

**Remark:** Note that, if all the intended receivers of packet *P* correctly receive it, there is no longer a packet-loss pattern relevant to packet *P*, vice versa. Accordingly, in the above example, the coded pattern $\rho_{P_1 \oplus P_2}$ still exits, even $R_1$ has already obtained packet $P_1 \oplus P_2$.

Furthermore, the coding node would retransmit *P* several times to deliver it successfully to its nexthop, for the reason that the medium is error-prone.

Clearly, there are no less than one lost packet relating to a loss pattern. Thereby, unless otherwise stated, we relate a loss pattern to a set of packets which have the same loss state. Then, we define $P_\rho$ as the set of lost packets which have the same loss pattern $\rho$ during the rescue process. Thus, in retransmission process, if a non-empty set $P_\rho$ is already transferred to the empty set, i.e. no loss packets relevant to $\rho$, we state that the coding node has *rescued* the loss pattern $\rho$ or $P_\rho$.

**Theorem 1.** *The expect number of retransmissions $\Omega_\rho$ requested by the central node to rescue loss-pattern $\rho$, whose the $i_r^{th}$ ($i_r \in N, r = 1, \ldots, N - W(\rho)$) entries are equal to 0, for N-receiver scenario is*

$$\Omega_\rho = \frac{|Or|}{1 - \prod \omega_{i_r}} \qquad (1)$$

*where Or is the set consisting of the lost packets which have the same loss pattern $\rho$ after the original transmission.. In particular, If and only if $W(\rho \oplus \rho') = W(\rho') - W(\rho) > 0$, $\Omega_{\rho \to \rho'}$ packets in $P_\rho$ would be transferred to $P_{\rho'}$, after the central node rescued $\rho$.*

$$\Omega_{\rho \to \rho'} = \Omega_\rho \times Pr\{\rho'\} / \prod \omega_{j_r} \quad (2)$$

where $j_r$ ($j_{r'} \in N, r' = 1, \ldots, W(\rho)$) are the entries in $\rho$ that are equal to 1. Particularly, if $\{j_r\} = \phi$, then Eq.(2) is transformed to

$$\Omega_{\rho \to \rho'} = \Omega_\rho \times Pr\{\rho'\} \quad (3)$$

**Proof:** To simplify the analysis, we suppose that node $C$ would retransmits $P_\rho$ *round by round*, which means node $C$ first transmits the entire packets belonging to $P_\rho$ one by one, and collects their receive-state to modify $P_\rho$, and then it repeats the these steps for the residual packets in $P_\rho$ again and again until these is no longer a packet relating to $\rho$. We define $P_\rho^{(k)}$ ($k > 0$) as the set relevant to $\rho$ after the $k^{th}$ round, and $P_\rho^{(0)} = Or$. Let random variable $Y_k$ ($k > 0$) represent the cardinality of $P_\rho^{(k)}$, and we have $Y_0 = |Or|$. For the reason that the deliveries are i.i.d. and follow the Bernoulli distribution, the random variables $Y_k$ ($k = 0, 1, \ldots$) are i.i.d. and follow the binomial distribution. Further-more, a packet is held in $P_\rho^{(k)}$ after the $k^{th}$ round, if and only if the receivers $R_{i_r}$ who lost it before are still failure to receive it. Hence, we have

$$E[Y_k] = E[E[Y_k|Y_{k-1}]] = E[Y_{k-1}] \times \prod \omega_{i_r}$$
$$= |Or| \times (\prod \omega_{i_r})^k \quad (4)$$

due to $\prod \omega_{i_k} \leqslant 1$, the series $E[Y_k]$ ($k > 0$) is convergent, the expect number of retransmissions $\Omega_\rho$, that the central node needs to rescues $\rho$, is

$$\Omega_\rho = Y_0 + \sum_{k=1}^{\infty} E[Y_k]$$
$$= Y_0 \cdot (Pr\{\rho\})^0 + \sum_{k=1}^{\infty} Y_0 \cdot (\prod \omega_{i_r})^k$$
$$= Y_0 \cdot \sum_{k=0}^{\infty} (\prod \omega_{i_r})^k = \frac{|Or|}{1 - \prod \omega_{i_r}} \quad (5)$$

As we mentioned above, the receivers who have obtained the packets belonging to $P_\rho$ before are no impact on the loss state $\rho$. Meanwhile, if a packet can be transferred from $P_\rho$ to $P_{\rho'}$, there must be $W(\rho') > W(\rho)$. It means that if the $q$ entry of $\rho$ is equal to 1, the corresponding entry of $\rho'$ must be 1. Consequently, we have $W(\rho \oplus \rho') = W(\rho') - W(\rho) > 0$ and the probability that a packet is transferred from $P_\rho$ to $P_{\rho'}$ is $Pr\{\rho'|\rho\} = Pr\{\rho'\} / \prod \omega_{j_k}$, particularly, if $\{j_r\} = \phi$, then $Pr\{\rho'|\rho\} = Pr\{\rho_2\}$. Now, let random variable $Y'_k$ ($k > 0$) represent the number of packets that are transferred from $P_\rho$ to $P_{\rho'}$, $\rho \neq \rho'$, after the $k^{th}$ round, particularly, $Y'_0 = 0$. Random variables $Y'_k$ ($k > 0$) are i.i.d as well and follow the binomial distribution. Then, we have

$$E[Y'_k] = E[E[Y'_k|Y_{k-1}]]$$
$$= E[Y_{k-1}] \times Pr\{\rho'|\rho\}$$
$$= Y_0 \times (\prod \omega_{i_r})^{k-1} \times Pr\{\rho'|\rho\} \quad (6)$$

due to $\prod \omega_{i_r} \leqslant 1$, the series $E[Y'_k]$ ($k > 0$) is convergent, the number of packets that are transferred from $P_\rho$ to $P_{\rho'}$ after node $C$ has rescued $\rho$ is Eq.

$$(7)$$

In particular, if $W(\rho) = N - 1$, there does not exit any $\rho'$

$$\begin{aligned}
\Omega_{\rho\to\rho'} &= \sum_{k=1}^{\infty} E[Y'_k] \\
&= \sum_{k=1}^{\infty} Y_0 \cdot (\prod \omega_{i_r})^{(k-1)} \cdot Pr\{\rho'|\rho\} \\
&= Pr\{\rho'|\rho\} \cdot Y_0 \cdot \sum_{k=0}^{\infty} (\prod \omega_{i_r})^k \\
&= \Omega_\rho \times Pr\{\rho'|\rho\}
\end{aligned} \qquad (7)$$

that packets in $P_\rho$ can be relocated to $P_{\rho'}$. ∎

Obviously, this theorem is suitable for both native and coded patterns. Finally, we introduce the following concept.

**Definition 4.** *If n packets respectively belonging to the sets $P_{\rho_1}, P_{\rho_2}, \ldots, P_{\rho_n}$ $(2 \leqslant n \leqslant N)$ can be coded together, we state the sets $P_{\rho_1}, P_{\rho_2}, \ldots, P_{\rho_n}$ can be encoded together, and we also say that the corresponding loss patterns $\rho_1, \ldots, \rho_n$ can be encoded together.*

## III. PERFORMANCE ANALYSIS

In this section, we carry out some theoretical analysis in terms of retransmission efficiency of the conventional and the proposed techniques for the wheel scenarios. Before delving into details, we refer the reader to the following symbols, which be used in the rest of the paper.

- $P^j$: The set of native packets node $R_j$ requests.
- $\rho^j$: The set of all possible loss patterns relating to the native packets node $R_j$ requests.
- $\rho^j_i$: The $i^{th}$ element in the set $\rho^j$.
- $\rho^{j_1,\ldots,j_n}$: The set of all possible coded patterns which are in relation to coded lost packets, each of which contains $n$ packets nodes $R_{j_1},\ldots,R_{j_n}$ requested separately, where $j_k \in N$, $1 \leqslant k \leqslant n$, $2 \leqslant n \leqslant N$.
- $\rho^{j_1,\ldots,j_n}_i$: The $i^{th}$ element in the set $\rho^{j_1,\ldots,j_n}$
- $Or^j_{\rho_k}$ is the set consisting of the lost packets that have the same loss pattern $\rho_k = \rho^j_k$ after the original transmission.

### A. NC-ARQ performs on the wheel topology

In this subsection, we study the performance of NC-ARQ technique in terms of the number of retransmissions over the wheel topologies as depicted in Fig. 1(a). We first analyze the "X" topology shown in Fig. 2.

*1) "X" topology:* There are two flows, $f_1(S_1 \to R_1)$ and $f_2(S_2 \to R_2)$ with the same deliver ratio which are twice as faster as node C. $R_1$ and $R_2$ is able to overhear $S_2$ and $S_1$ separately. For the sake of simplicity, we assume that the loss rate of each snoopy links is zero, i.e. as long as a node enables to listen to other nodes, it can always successfully overhear packets from them. Because of the error-prone charater of wireless channels, $S_1$(or $S_2$) would deliver a packet several times to correctly transmit it to node $C$, i.e. its neighbor $R_1$(or $R_2$) has a relatively high probability to overhear this packet. Hence, the above assumption is not crucial to our analysis and is always valid in the rest of this paper. Using the assumption, if there are two different packets that came from $S_1$ and $S_2$ respectively, node C will mix them together to one packet and broadcast it to $R_1$ and $R_2$. And based on the results of Ref. [5], C would send out the encoded packets all the time in original transmission, when overall throughput is stable.

Now let us focus on the retransmissions between node C and its neighbors. Fig. 2(b) lists all possible loss patterns. Since each receiver enable to overhear the entire packets which are sent by its neighbor, coded pattern $\rho^{1,2}_1$ is actually equivalent to the native pattern $\rho^1_1 = [0\ 1]$, and $\rho^{1,2}_2$ is corresponding $\rho^2_1 = [1\ 0]$. Apparently, these two patterns can be XORed together, yet lost packets relevant to $\rho^{1,2}_0$ would be retransmitted by the same form. Using the NC-ARQ, we have the following result,

**Theorem 2.** *Using the NC-ARQ technique, when the number of packets to be sent is sufficiently*

large, the average number of retransmissions $\lambda_{XNC}$ for the "X" scenario is

$$\lambda_{XNC} = \frac{1}{2} \times \frac{max\{\omega_1, \omega_2\}}{1 - max\{\omega_1, \omega_2\}} \tag{8}$$

The proof of the theorem is based on the method in [9]. In the proof, we need the following lemma

**Lemma 1.** *Using the NC-ARQ technique, for N-receiver scenario, if $\rho_i^{r_1}$ and $\rho_j^{r_2}$ can be coded together and $|P_{\rho_i}^{r_1}| \leqslant |P_{\rho_j}^{r_2}|$, $\omega_{r_1} \leqslant \omega_{r_2}$, then the entire packets in $P_{\rho_i}^{r_1}$ can be XORed with the ones belonging to $P_{\rho_j}^{r_2}$ in recovery process, and we say that $\rho_i^{r_1}$ is dominated by $\rho_j^{r_2}$. This relation is denoted by $\rho_i^{r_1} \succ \rho_j^{r_2}$. In particular, node C need to retransmit $\Omega_{\rho_i}^{r_1}$ times to rescue $\rho_i^{r_1}$. And $\Omega_{Native}$ native packets would be sent alone.*

$$\Omega_{\rho_i}^{r_1} = \frac{|P_{\rho_i}^{r_1}|}{1 - \omega_{r_1}} \tag{9}$$

$$\Omega_{Native} = |P_{\rho_j}^{r_2}| - \frac{|P_{\rho_i}^{r_1}|(1 - \omega_{r_2})}{1 - \omega_{r_1}} \tag{10}$$

We use induction method to prove the lemma. Interested readers can find details of the proof in the Appendix.

**Proof of theorem 2:** We assume that node C wants to transmit K packets to each receiver. As we described earlier, node C would transmit K coded packets in the original transmission. Without loss of generality, we suppose that the packet loss ratio $\omega_1 \leqslant \omega_2$. Let random variable $X_0$, $X_1$ and $X_2$ separately denote the number of lost packets relevant to loss pattern $\rho_0^{1,2}$, $\rho_1^{1,2}$ and $\rho_2^{1,2}$ after K transmissions. $X_k$ ($k = 0, 1, 2$) follow the binomial distribution. Then, we have

$$E[X_k] = K \Pr\{\rho_k^{1,2}\}, \ k = 0, 1, 2 \tag{11}$$

where $Pr\{\rho_k^{1,2}\}, k = 0, 1, 2$ are the probabilities that the loss patterns $\rho_k^{1,2}, k = 0, 1, 2$ happen, respectively.

Because of $\omega_1 \leqslant \omega_2$, then we get $\Pr\{\rho_1^{1,2}\} \leqslant \Pr\{\rho_2^{1,2}\}$. Thus, the expect number of loss pattern $\rho_2^{1,2}$ is no less than $\rho_1^{1,2}$, $E[X_1] \leqslant E[X_2]$. As we mentioned above, when node C rescues $\rho_0^{1,2}$, some packets would be transfer from $P_{\rho_0}^{1,2}$ to $P_{\rho_1}^{1,2}$ or $P_{\rho_2}^{1,2}$. And packets belonging to $P_{\rho_1}^{1,2}$ or $P_{\rho_2}^{1,2}$ is unable to be relocated to $P_{\rho_0}^{1,2}$. Hence, node C would rescue $\rho_0^1$ first. Then based on Theorem 1, we have

$$\Omega_{\rho_0^{1,2}} = \frac{E[X_0]}{1 - \omega_1 \omega_2} \tag{12}$$

$$\Omega_{\rho_0^{1,2} \to \rho_1^{1,2}} = \Omega_{\rho_0^{1,2}} \cdot \omega_1(1 - \omega_2) \tag{13}$$

$$\Omega_{\rho_0^{1,2} \to \rho_2^{1,2}} = \Omega_{\rho_0^{1,2}} \cdot (1 - \omega_1)\omega_2 \tag{14}$$

Clearly $\Omega_{\rho_0^{1,2} \to \rho_1^{1,2}} \leqslant \Omega_{\rho_0^{1,2} \to \rho_2^{1,2}}$. Hence, $E[X_1] + \Omega_{\rho_0^{1,2} \to \rho_1^{1,2}} = m_1 \leqslant m_2 = E[X_2] + \Omega_{\rho_0^{1,2} \to \rho_2^{1,2}}$. Based on Lemma 1, we directly get $\rho_1^{1,2} \succ \rho_2^{1,2}$, and

$$\Omega_{\rho_1^{1,2}} = \frac{m_1}{1 - \omega_1} \tag{15}$$

$$\Omega_{Native} = m_2 - \frac{m_1(1 - \omega_2)}{1 - \omega_1} \tag{16}$$

Then, let random variable Y denote the number of retransmissions needed to get one success at $R_2$. And Y follows the geometric distribution, $E[Y] = \frac{1}{1-\omega_2}$. at receiver $R_2$. To sum up, the expected number of retransmissions to success-fully deliver K coded packets to $R_1$ and $R_2$ is given by

$$\Omega = \Omega_{\rho_0^{1,2}} + \Omega_{\rho_1^{1,2}} + E[Y]\Omega_{Native} = \frac{K\omega_2}{1-\omega_2} \tag{17}$$

Dividing 2*K* by $\Omega$, then gives the theorem. ∎

**Remark:** Note that in this case, to recover all loss packets is equivalent to rescue the loss packets relating to the receiver who has the higher packet loss rate, namely a receiver with better link-state is dominated by the one with worse link-state.

*2) The wheel topology*: Due to the fact that the benefit of encoded packets correspond to the irrelevant flows, we only focus on the wheel topology, which consists of *N* receivers and $N-2$ irrelevant flows, in this paper. We will address the impact of the relevant flows on retransmission efficiency in future work.

**Theorem 3.** *Using the NC-ARQ technique, when the number of packets to be sent is sufficiently large, the expect retransmissions efficiency $\lambda_{WNC}^N$ for the N-receiver and $N-2$ irrelevant flows wheel topology is .*

$$\lambda_{WNC}^N = \frac{1}{N}\left(\frac{\omega_{r_1}}{1-\omega_{r_1}} + \sum_{i=1,i\neq r_1,r_2}^{N} \frac{\prod_{j=i,j\neq r_1,r_2}^{N}\omega_j}{1-\omega_i} \right.$$
$$+ \frac{(\omega_{r_2}-\omega_{r_1})(1-\omega_{r_2})}{1-\omega_{r_1}} \sum_{i=1,i\neq r_1}^{r_2-1} \frac{\prod_{j=i,j\neq r_1,r_2}^{N}\omega_j}{1-\omega_i} \tag{18}$$
$$\left.+ \frac{(\omega_{r_2}-\omega_{r_1})(\prod_{j=r_2}^{N}\omega_j)}{(1-\omega_{r_1})(1-\omega_{r_2})}\right)$$

where $\omega_j$ is the packet loss ratio between coding node and receiver $R_j$, and $\omega_i \leqslant \omega_j$ if $1 \leqslant i \leqslant j \leqslant N$. $R_{r_1}$ and $R_{r_2}$ relate to the relevant flows, $r_1 < r_2$ and $r_1, r_2 \in N$

**Proof:** Using the NC-ARQ technique, node *C* is unable to combine any other packet(s) with the coded packet as $P_s^{r_1} + P_t^{r_2}$. Therefore, we suppose that node *C* first transmits $P_s^{r_1} + P_t^{r_2}$ correctly to $R_{r_1}$ and $R_{r_2}$. Based on Theorem 2, we immediately obtain the number of such coded packets node *C* retransmits in all

$$\Omega_{Code} = \frac{K\omega_{r_1}\omega_{r_2}}{1-\omega_{r_1}\omega_{r_2}} + \frac{K\omega_{r_1}(1-\omega_{r_2})}{(1-\omega_{r_1})(1-\omega_{r_1}\omega_{r_2})}$$
$$= \frac{K\omega_{r_1}}{1-\omega_{r_1}} \tag{19}$$

and there are also $\Omega_{Native}$ packets *C* needs to deliver to $R_{r_2}$.

$$\Omega_{Native} = \frac{K(\omega_{r_2}-\omega_{r_1})}{1-\omega_{r_1}} \tag{20}$$

Thereafter, node *C* begins to retransmit the rest $(N-2)K + \Omega_{Native}$ packets to their intended destinations, respectively. Obviously, this scenario is the same as the one-to-many single-hop unicast scenario we described earlier. Note that $\Omega_{Native} \leqslant K$, therefore, we further separate node *C*'s transmission into two independent procedures, first node *C* sends $(N-1)\Omega_{Native}$ packets correctly to the receivers $R_i$ $(i=1,\ldots,N, i\neq r_1)$; then *C* transmits the rest $(N-2)$ to the receivers $R_i$ $(i=1,\ldots,N, i\neq r_1,r_2)$. Based on Theorem 4.3 in [9], we calculate the total expect number of retransmissions of these two procedures, respectively,

$$\Omega_1 = \Omega_{Native} \cdot \sum_{i=1,i\neq r_1}^{N} \frac{\prod_{j=i,j\neq r_1}^{N}\omega_j}{1-\omega_i} \tag{21}$$

$$\Omega_2 = (K - \Omega_{Native}) \cdot \sum_{i=1,i\neq r_1,r_2}^{N} \frac{\prod_{j=i,j\neq r_1,r_2}^{N}\omega_j}{1-\omega_i} \tag{22}$$

Combining Eq. (19), (21) and (22) together and dividing $KN$, then gives the theorem. ∎

## B. The proposed approach performs on the wheel topology

In this subsection, we evaluate the performance of the proposed approach for the wheel topology. We first present how to use packet-loss pattern to identify the coding opportunities over retransmission process.

**Theorem 4.** Let $R^{\rho_i}$ denote the set consisting of node(s) which is/are the intended receiver(s) for packets in $P_{\rho_i}$. In rescue process, the loss packets $P_1, P_2,...,P_n$ $(2 \leqslant n \leqslant N)$, each of which can be native or coded packet, with respect to the patterns $\rho_1,\ldots,\rho_n$ can be coded together if and only if $R^{\rho_i} \cap R^{\rho_j} = \phi, i \neq j$ and only the $i^{th}$ entry of the $j_k^{th}$ column of matrix $F = [\rho_1^T \ \ldots \ \rho_n^T]^T$ is equal to 0, where $R_{j_k} \in R^{\rho_i}$, $i \in [1,n]$ and $k \in [1, |\bigcup_{i=1}^{n} R^{\rho_i}|]$.

**Proof:** First of all, let us assume that there are two loss patterns $\rho_{i_1}$ and $\rho_{i_2}(i_1 \neq i_2, i_1, i_2 \in n)$ which satisfy $R^{\rho_{i_1}} \cap R^{\rho_{i_2}} = \{R^{cap}\}$. Suppose $P_{i_1}$ contains the native packet $P_1^{cap}$ and the corresponding packet in $P_{i_2}$ is $P_2^{cap}$. Under these assumptions, there would be two possibilities: 1. $P_1^{cap} \neq P_2^{cap}$, obviously, $R^{cap}$ cannot decode such coded packet; 2. $P_1^{cap} = P_2^{cap}$, then $R^{cap}$ cannot obtain this packet, for the reason $P_1^{cap} \oplus P_2^{cap} = 0$. Neither of them is expected by us. Hence, there must be $R^{\rho_i} \cap R^{\rho_j} = \phi, i \neq j$.

Without loss of generality, let us consider node $R_{j_1}$. To insure that the node $R_{j_1}$ get its corresponding packet $P_1$, it must have correctly overheard the packets $P_i, i = 2,\ldots,n$. So based on Definition 2, the $j_1^{th}$ entry in $\rho_i, i = 2,\ldots,n$ is equal to 1, i.e. the $j_1^{th}$ column of matrix $F$ is only the first entry that is equal to 0. It is the same to other receivers. On the other hand, if the $j_k^{th}$ column of matrix $F$ is only the $i^{th}$ entry that is equal to 0, where $R_{j_k} \in R^{\rho_i}$. It means that $R_{j_k}$ has already obtained the packets $P_t, t = 1,\ldots,n, t \neq i$, i.e. it enable to eliminate $P_t$ from the coded packet. Under the assumption, all packets except $P_i$ can be removed by $R_{j_k}$. Clearly, there are the same results for all other receivers. Therefore, based on the code rule of COPE [10], all the packets can be mixed together to one packet. ∎

**Definition 5.** The set
$$CG = \{\rho_1, \rho_2, \ldots, \rho_k\}, 2 \leqslant k \leqslant N \quad (23)$$
where error patterns $\rho_1, \rho_2, \ldots, \rho_k$ can be coded together, is called the Code Group.

Particularly, though there are no less than one code group interrelating with a error pattern, if all the $j_k^{th}$ ($j_k \notin \bigcup_{i=1}^{n} R^{\rho_i}$) columns of matrix $F$ must be zero in Theorem 4, and then each combination of loss patterns fulfilling the theorem is unique, for example if $\rho_i \in CG$ and $\rho_i \in CG'$, then $CG = CG'$, and achieve their maximum coding opportunities. In this way, all the code groups we mentioned are limited to the restriction in the rest of this paper.

Let us now focus on the 3-receiver wheel topology as shown in Fig. 1(a) and suppose $\omega_1 \leqslant \omega_2 \leqslant \omega_3$. Table in Fig. 4(a) lists all possible corresponding loss patterns. Based on Theorem 5, $\rho_0^{1,2}$, $\rho_1^{1,2}$ and $\rho_0^{1,2}$ cannot be coded with any loss pattern in $\rho^3$. And as we discussed above, $R_2$ will dominate the rescue process of $\rho_0^{1,2}$, $\rho_1^{1,2}$ and $\rho_0^{1,2}$. Moreover, if C enables to be conscious of encoded packets $R_3$ overheard, then we observe that $\rho_4^{1,2}$ could be mixed with $\rho_2^3$, however, the best combination is to combine $\rho_4^{1,2}$, $\rho_5^{1,2}$ and $\rho_3^3$ together, because such one includes more information. We can easily prove $\{\rho_4^{1,2}, \rho_5^{1,2}\} \prec \rho_3^3$, i.e. the entire packets in $P_{\rho_4}^{1,2}$ and $P_{\rho_5}^{1,2}$ can be XORed with the ones in $P_{\rho_3}^3$. It means that all the packets in $P_{\rho_0}^3, P_{\rho_1}^3, P_{\rho_2}^3$ and the partial in $P_{\rho_3}^{1,2}$ would be transmitted alone, sine none of them can be coded together based on Theorem 5.

Nevertheless, we first let node $C$ combine $\rho_3^{1,2}$ with $\rho_1^3$, and then combine $\rho_3^{1,2}$ with $\rho_2^3$ again, namely we separate $P_{\rho_3}^{1,2}$ into two uniform sets $P_{\rho_3}^1$ and $P_{\rho_3}^2$, where $P_{\rho_3}^1 = P_{\rho_3}^2 = P_{\rho_3}^{1,2}$ and $\rho_3^1 = \rho_3^2 = \rho_3^{1,2}$, and rescue them respectively. Then an interesting thing would be observed, that although $C$ would free $P_{\rho_3}^{1,2}$ at twice, the total number of retransmissions decreases. So, because of $\rho_3^1 \prec \rho_2^3$, the flow $f_1(S_1 \to R_1)$ would be "absorbed" by $f_2(S_2 \to R_2)$ and $f_3(S_3 \to R_3)$, since the entire packets in $P^1$ are XORed with $P^2$ and $P^3$ (clearly it is valid in original transmission). Thus, $R_1$ and $R_2$ can be regarded as one node, i.e. the 3-receiver wheel topology transform to the 2-receiver single-hop unicast scenario.

Now, in the light of the front illustration, we summarize the following redistribution algorithm in Fig. 5 for N-receiver two-relevant-flow wheel topology, where node $R_{r_1}$ and $R_{r_2}$ relate to the relevant flows, $r_1, r_2 \in N$, $\omega_{r_1} \leqslant \omega_{r_2}$.

Using the this algorithm, we have the following theorem.

**Theorem 5.** Let $f_{r_1}$ and $f_{r_2}$ denote the relevant flows in N-receiver 2-relevant-flow wheel scenario, where $r_1, r_2 \in N$ and $\omega_{r_1} \leqslant \omega_{r_2}$. Using the proposed algorithm, if the number of packets to be sent is sufficiently large, then the flow $f_{r_1}$ is able to be absorbed by $f_{r_2}$ and the irrelevant flows, in other words this scenario is equivalent to $N-1$ receivers unicast scenario. The average number of retransmissions $\lambda_{WNC}^E$ is

$$\lambda_{WNC}^E = \frac{1}{N} \sum_{i=1, i \neq r_1}^{N} \frac{\prod_{j=i, j \neq r_1}^{N} \omega_j}{1 - \omega_i} \qquad (24)$$

In the proof, we need the following concepts

**Lemma 2.** Let $CG = \{\rho_{i_1}^{r_1}, \ldots, \rho_{i_n}^{r_n}\}$, where $r_n \in N$, $i_n \in [0, |\rho^{r_n}|]$. If $|P_{\rho_{i_k}}^{r_k}| \leqslant |P_{\rho_{i_j}}^{r_j}|$ and $\omega_{r_k} \leqslant \omega_{r_j}$, $r_k \leqslant r_j$, then all the packets in $P_{\rho_{i_j}}^{r_j}$, $j = 1, \ldots, n-1$ can be XORed with the ones belonging to $P_{\rho_{i_n}}^{r_n}$ over rescue process, thereby we say that $CG$ is dominated by $\rho_{i_n}^{r_n}$. This relation is denoted by $CG \succ \rho_{i_n}^{r_n}$.

**Proof:** Let random variable $X_k$ ($1 \leqslant k \leqslant n$) respectively denote the number of retransmissions needed to successfully deliver all lost packets relating to $\rho_{i_{k'}}^{r_k}$. Let random variable $Y_k$ ($1 \leqslant k \leqslant n$) denote the number of tries before a effective transmission for user $R_{r_k}$, separately. $Y_k$ follows the geometric distribution, $E[Y_k] = \frac{1}{1-\omega_{r_k}}$, and $E[X_k] = |Or_{\rho_{i_k}}^{r_k}| \cdot E[Y_k]$. Because of the assumption, $\rho_{i_k}^{r_k}$ and $\rho_{i_j}^{r_j}$ can be XORed together and $E[X_k] \leqslant E[X_j]$ ($k \leqslant j$). Therefore, after $E[X_j]$ transmissions, pattern $\rho_{i_k}^{r_k}$ has already been rescued, for the reason that all packets in $P_{\rho_{i_k}}^{r_k}$ has been XORed with the ones belonging to $P_{\rho_{i_j}}^{r_j}$. Then, this relation is called $\rho_{i_j}^{r_j}$ dominates $\rho_{i_k}^{r_k}$, $\rho_{i_k}^{r_k} \succ \rho_{i_j}^{r_j}$. Carrying the same argument through, we can prove that $\rho_{i_n}^{r_n}$ dominates any other loss-pattern in $CG$, which means that to rescue $CG$ is equivalent to recovering $\rho_{i_n}^{r_n}$, in other words $CG$ is dominated by $\rho_{i_n}^{r_n}$.

In particular, after $E[X_{n-1}]$ deliveries, there is only $P_{\rho_{i_n}}^{r_n} \neq \phi$, namely there are

$$\Omega_{Native} = (E[X_n] - E[X_{n-1}])/E[Y_k]$$
$$= \left|Or_{\rho_{i_n}}^{r_n}\right| - \frac{\left|Or_{\rho_{i_{n-1}}}^{r_{n-1}}\right|(1 - \omega_{r_n})}{1 - \omega_{r_{n-1}}} \qquad (25)$$

native packets should be transmitted alone. ∎

**Proof of Theorem 5:** Obviously, flow $f_{r_1}$ is dominated by flow $f_{r_2}$ in original transmission. In retransmission process, after redistributing, the loss packets relating to node $r_1$ are separated to two disjoint sets: one contains the packets having non-zero error pattern, but none of $r_1$ and $r_2$ received; the residual packets belong to the other one. The former one associates with $\rho^{r_1}$, and the later one with $\rho^{r_1+r_2}$.

For $\rho_i^{r_1} \in \rho^{r_1}$, $i \in [1, |\rho^{r_1}|]$, there are only $W(\rho_i^{r_1})$ patterns $\rho_{i_1}^{j_1}, \rho_{i_2}^{j_2}, \ldots, \rho_{i_W}^{j_W}$, where $W = W(\rho_i^{r_1})$ and $W(\rho_{j_k}^{i_k}) = W(\rho_i^{r_1})$, $k \in [1, W]$ and it's well-determined for each $\rho_i^{r_1}$, that can be combined with $\rho_i^{r_1}$. If $\rho_i^{r_1}$ is dominated by one of them, then the loss packets in $P_{\rho_i}^{r_1}$ would incorporate into flow to which it belongs. If $\rho_i^{r_1}$ dominates the coded group, then based on Lemma 2 there are $\Omega_N^{r_1}$ native packets in $P_{\rho_i}^{r_1}$ that have to be deliver alone. We suppose the sub-dominant pattern is $\rho_{j_n}^{i_n}$, $n \in [1, W]$, so

$$\Omega_N^{r_1} = \left|P_{\rho_i}^{r_1}\right| - \frac{\left|P_{\rho_{i_n}}^{j_n}\right| \cdot (1 - \omega_{r_1})}{1 - \omega_n}$$

Note that $\rho_i^{r_2}$ ($\rho_i^{r_2} = \rho_i^{r_1}$) also combines $\rho_{i'_1}^{j_1}, \rho_{i'_2}^{j_2}, \ldots, \rho_{i'_W}^{j_W}$, where the $r_1^{th}$ and $r_2^{th}$ entries of $\rho_{i_k}^{j_k} \oplus \rho_{i'_k}^{j_k}$ are equal to 1, and $W(\rho_{i_k}^{j_k} \oplus \rho_{i'_k}^{j_k}) = 2$. In this scenario, the corresponding sub-dominant pattern is $\rho_{i'_n}^{j_n}$ and we have

$$\Omega_N^{r_2} = \left|P_{\rho_i}^{r_2}\right| - \frac{\left|P_{\rho_{i'_n}}^{j_n}\right| \cdot (1 - \omega_{r_2})}{1 - \omega_n}$$

Because of $\omega_{r_1} \leqslant \omega_{r_2}$, we have $\Omega_N^{r_1} \leqslant \Omega_N^{r_2}$. And then, in the light of Lemma 1, $\rho_i^{r_1}$ is always dominated by $\rho_j^n$ and $\rho_i^{r_2}$. On the other hand, for $\rho_i^{r_1+r_2} \in \rho^{r_1+r_2}$, $i \in [0, |\rho^{r_1+r_2}|]$, if it relates to $r_1$, a unique pattern $\rho_{i'}^{r_1+r_2}$ relating to $r_2$ and satisfying the $r_1^{th}$ and $r_2^{th}$ entries of $\rho_{i'}^{r_1+r_2} \oplus \rho_i^{r_1+r_2}$ are equal to 1, $W(\rho_{i'}^{r_1+r_2} \oplus \rho_i^{r_1+r_2}) = 2$ will be found. Due to $\omega_{r_1} \leqslant \omega_{r_2}$, we have $|P_{\rho_i}^{r_1+r_2}| \leqslant P_{\rho_{i'}}^{r_1+r_2}$, so $\rho_i^{r_1+r_2} \succ \rho_{i'}^{r_1+r_2}$.

In short, $f_{r_1}$ can be absorbed by $f_{r_2}$ and other flows. Based on Theorem 4.3 in [9], we directly calculate the expect retransmissions efficiency is equal to Eq. (24). ■

## IV. EXPERIMENTAL RESULTS AND DISCUSSIONS

In this section, retransmission gain is used to evaluate the retransmission efficiency of different approaches by varying the number of receivers and bit error rate (BER) under both unicast and wheel topology. The BER at each receiver are mutually independent and follow the Bernoulli distribution. We define the retransmission gain as the total number of retransmissions using a typical retransmission algorithm, which is the HARQ or NC-HARQ techniques, divided by the total number of retransmissions using our algorithm. A higher retransmission gain is preferred since it indicates fewer retransmissions. In the following simulate, the packet size is to be 1532 bytes and data is encoded with RS (32, 28, 4). We use CRC-16 for error detection in all the simulations. We record the total number of retransmissions over a mass of experiments. In the interest of space and clarity, we only present the average retransmission gains in the following simulation results.

First, we compare the performance of proposed approach with the NC-HARQ technique and the HARQ technique. Fig. 5 show the effect of different BERs on the retransmission gain for the 3-receiver wheel topology. BERs between the sender and all its receivers are the same, and varied from $10^{-4}$ to $3.5 \times 10^{-3}$ in $5 \times 10^{-4}$ increments. As expected, the simulation results support our theoretical derivations. Furthermore, an interesting phenomenon would be observed is that the retransmission gains for unicast scenario increase with BER, and are very close to 1 under the light BERs. The reason is that the more awful noise the channels encounter, the more packets are lost, that is to say the more encoded packets would be generated from which the proposed protocol enables to explore the coding opportunities, yet NC-ARQ cannot. On the contrary, the performance gains for wheel topology decrease from 2 to 1 as BER rises. This is due to the fact that when BERs rises, the proportion of loss packets relating to $|P_{\rho_3}^{1+2}|$ falls off, then the superiority of the proposed approach over NC-ARQ, in turn, becomes gradually less.

Then, we compare the performance of proposed approach with the NC-HARQ technique versus the number of clients for unicast and wheel topology. In the experiment, BERs between the sender and all its receivers are set equal to $2 \times 10^{-3}$ and $3 \times 10^{-3}$. Fig. 6 show the impact of different number of receivers, which is varied from 3 to 25, on the retransmission gain. As seen, the retransmission gains hardly increase with the number of clients. The reason is that the probability that a loss packet is successfully received at lease at one client increases with the number of the clients, and then more loss packets relating to one namely the growth rate of the total number of retransmissions is getting slower and slower, when the number of receivers goes up.

## V. CONCLUTIONS AND FUTURE WORK

In this paper, we addressed the problem of existing NC-based reliable transmission schemes for wireless unicast system such as WiFi and WiMAX networks. In order to overcome the disregard for the role of encoded packets, an efficient rescue approach based on XOR network coding has been proposed. The core feature of the proposed approach is that the coding node explores the coding opportunities from encoded packet to achieve the optimal coding decisions. Under the access point model, some analytical results are derived for the retransmission efficiency over the wheel scenario. Further-more, the theoretical and simulation results indicate that the proposed approach always outperforms the traditional HARQ and the NC-HARQ technique in terms of the retransmission efficiency.

APPENDIX

**Proof of lemma 1:** We still suppose that the coding node retransmits $P_{\rho_i}^{r_1}$ round by round. We define $P_{\rho_i}^{r_1}(k)$ and $P_{\rho_j}^{r_2}(k)$ $(k>0)$ as the loss packets sets relevant to $\rho_i^{r_1}$ and $\rho_j^{r_2}$ after the $k^{th}$ round and we get $P_{\rho_i}^{r_1}(0) = P_{\rho_i}^{r_1}, P_{\rho_j}^{r_2}(0) = P_{\rho_j}^{r_2}$. Let random variable $Y_k$ and $Z_k(k>0)$ represent the cardinality of $P_{\rho_i}^{r_1}(k)$ and $P_{\rho_j}^{r_2}(k)$, furthermore, we define $Y_0 = |P_{\rho_i}^{r_1}|, Z_0 = |P_{\rho_j}^{r_2}|$. As the deliveries are i.i.d. and follow the Bernoulli distribution, random variables $Y_k$ and $Z_k$ $(k = 0, 1, \dots)$ are i.i.d too and follow the Binomial distribution. Due to $Y_0 \leqslant Z_0$, the entire packets in $P_{\rho_i}^{r_1}(0)$ is able to be combined with the ones belonging to $P_{\rho_j}^{r_2}(0)$ in next round. After the first round, $Y_1 = \omega_{r_1} Y_0$ and $m_0 = Z_0 - Y_0$ packets in $P_{\rho_j}^{r_2}(0)$ are not delivered, i.e. $Z_1 = \omega_{r_2} Y_0 + m_0$. Because of $\omega_{r_1} \leqslant \omega_{r_2}$, we explicitly have $Y_1 \leqslant Z_1$. It means that the entire packets within $P_{\rho_i}^{r_1}(1)$ can be mixed with the ones belonging to $P_{\rho_j}^{r_2}(1)$ over the next round in the same way.

Now, we suppose that $Y_n \leqslant Z_n$ $(n \geqslant 3)$. Then after the $n^{th}$ round, node $R_1$ is failure to receive $Y_{n+1} = \omega_{r_1} Y_n$ packets, so the coding node has to retransmit these packets in the next round. Moreover, $m_n = Z_n - Y_n$ packets in $P_{\rho_j}^{r_2}(n)$ are not delivered, i.e. $Z_{n+1} = \omega_{r_2} Y_n + m_n$. Clearly, we have $Y_{n+1} \leqslant Z_{n+1}$, which indicates that we are able to combine the entire packets in $P_{\rho_i}^{r_1}(n+1)$ with the ones belonging to $P_{\rho_j}^{r_2}(n+1)$. Then we deduce $\rho_1 \succ \rho_2$ by induction, and work out the expect number of retransmissions required to rescue $\rho_1$.

$$\Omega'_{Code} = \sum_{k=0}^{\infty} Y_k = \frac{|P_{\rho_1}|}{1 - \omega_{r_1}} \tag{26}$$

In particular, $\Omega'_{Native}$ native packets relevant to $\rho_2$ would be delivered alone.

$$\begin{aligned}\Omega'_{Native} &= Z_\infty \\ &= Z_0 + \sum_{k=0}^{\infty} (\omega_{r_2} - 1) Y_k \\ &= |P_{\rho_2}| - \frac{|P_{\rho_1}|(1 - \omega_{r_2})}{1 - \omega_{r_1}}\end{aligned} \tag{27}$$

The lemma has been proved. ∎

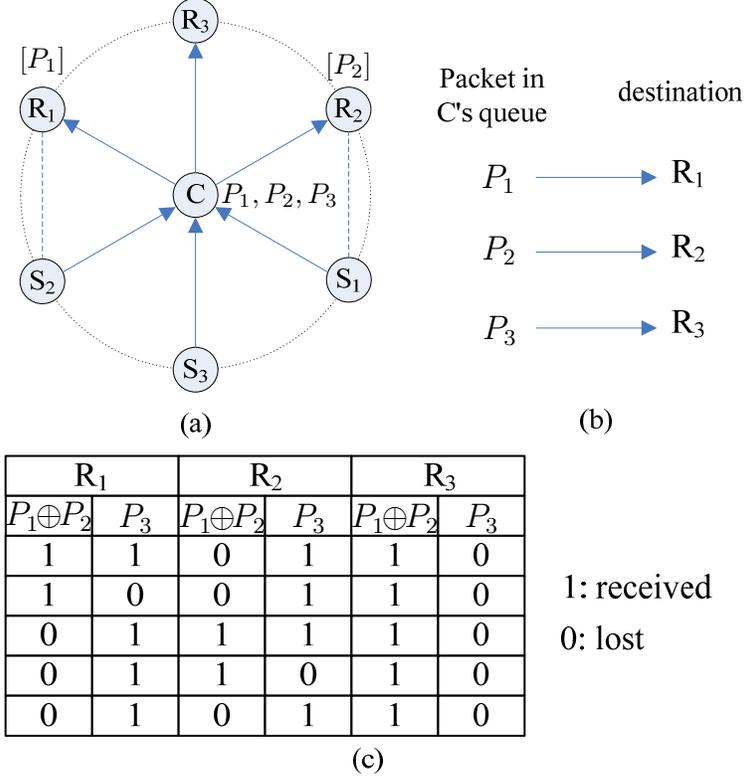

Figure 1. Example of the coding opportunities of encoded packets. (a) A wheel topology with three flows. (b) Nexthops of packets in C's queue. (c) Possible packet-loss states.

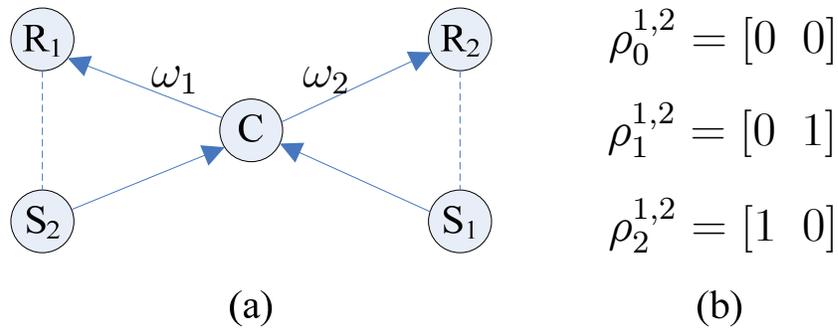

Figure 2. (a) The "X" scenario. (b) All possible packet-loss patterns.

| $\rho^{1,2}$ | $R_1$ | $R_2$ | $R_3$ |
|---|---|---|---|
| $\rho_0^{1,2}$ | 0 | 0 | 0 |
| $\rho_1^{1,2}$ | 0 | 1 | 0 |
| $\rho_2^{1,2}$ | 1 | 0 | 0 |
| $\rho_3^{1,2}$ | 0 | 0 | 1 |
| $\rho_4^{1,2}$ | 0 | 1 | 1 |
| $\rho_5^{1,2}$ | 1 | 0 | 1 |

| $\rho^3$ | $R_1$ | $R_2$ | $R_3$ |
|---|---|---|---|
| $\rho_0^3$ | 0 | 0 | 0 |
| $\rho_1^3$ | 0 | 1 | 0 |
| $\rho_2^3$ | 1 | 0 | 0 |
| $\rho_3^3$ | 1 | 1 | 0 |

Figure 3.  All possible loss patterns for 3-receiver wheel topology.

---

Redistribution algorithm:
1: **for** $i = 1$ to $|\rho^{r_1,r_2}|$ **do**
2: Pick $\rho_i^{r_1,r_2}$ from $\rho^{r_1,r_2}$
3: **if** the $r_1^{th}$, $r_2^{th}$ entries of $\rho_i^{r_1,r_2}$ are 0, and $W(\rho_i^{r_1,r_2}) \neq 0$ **then**
4: $\rho_{|\rho^{r_1}|+1}^{r_1} = \rho_{|\rho^{r_2}|+1}^{r_2} = \rho_i^{r_1,r_2}$
5: $P_{\rho_{|\rho^{r_1}|+1}}^{r_1} = P_{\rho_{|\rho^{r_2}|+1}}^{r_2} = P_{\rho_i}^{r_1,r_2}$
6: delete $\rho_i^{r_1,r_2}$ and $P_{\rho_i}^{r_1,r_2}$
7: **end if**
8: **end for**

---

Figure 4.  Redistribution algorithm for the wheel topology

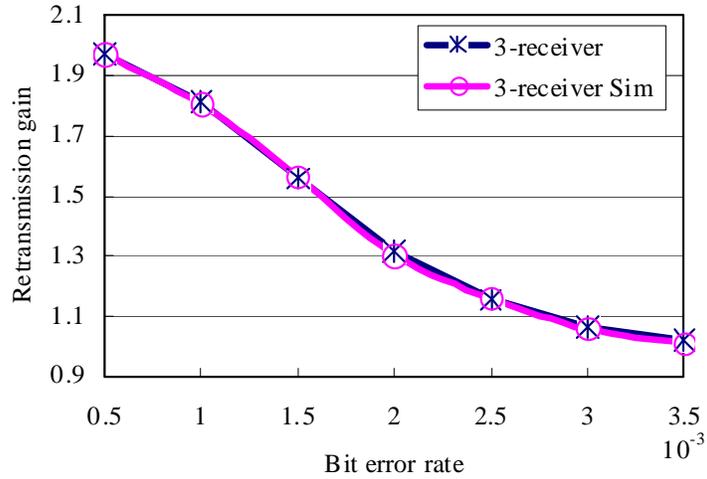

Figure 5.  retransmission gain versus BER for theory and simulation.

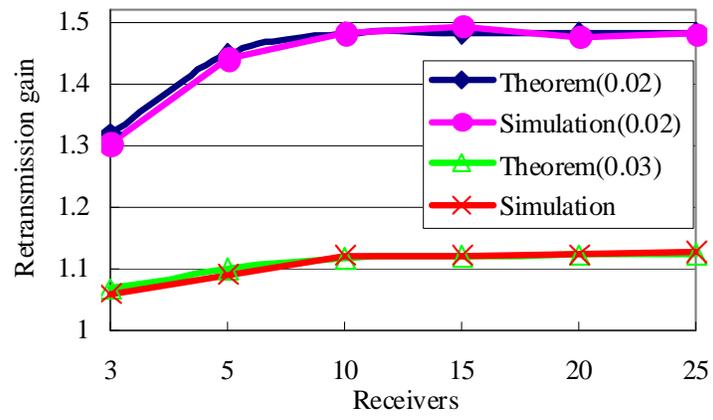

Figure 6. retransmission gain versus the number of receivers for theory and simulation.